\documentclass[sigconf]{acmart}

\acmBooktitle{Proceedings of the CHI 2026 Workshop: Ethics at the Front-End: Responsible User-Facing Design for AI Systems (CHI '26), April 13--17, 2026, Barcelona, Spain}

\usepackage{comment}

\AtBeginDocument{%
  }

\newtoggle{comments}
\toggletrue{comments}
\iftoggle{comments} {
}

\setcopyright{acmlicensed}
\copyrightyear{2026}
\acmYear{2026}
\acmConference[CHI '26]{ACM Conference on Human Fact CHI conference on Human Factors in Computing}{April 13--17, 2026}{Barcelona, Spain}
\begin{document}

%%
%% The "title" command has an optional parameter,
%% allowing the author to define a "short title" to be used in page headers.
\title{Interpretative Interfaces: Designing for AI-Mediated Reading Practices and the Knowledge Commons}
%Interpretative Interfaces: \\ Reframing AI Interpretability as Interface Design}
%other options: "Interpretative Interfaces: Designing for Critical Reading in AI-Mediated Scholarship"
% "Interpretative Interfaces for AI-Mediated Reading: Reframing Interpretability as Design"
% "Reading Between the Layers: Interpretative Interface Design for LLM-Mediated Scholarship"

%%
%% The "author" command and its associated commands are used to define
%% the authors and their affiliations.
%% Of note is the shared affiliation of the first two authors, and the
%% "authornote" and "authornotemark" commands
%% used to denote shared contribution to the research.
\author{Gabrielle Benabdallah}
\email{gabben@uw.edu}
\affiliation{%
  \institution{University of Washington}
  \city{Seattle}
  \state{Washington}
  \country{USA}
}

%%
%% By default, the full list of authors will be used in the page
%% headers. Often, this list is too long, and will overlap
%% other information printed in the page headers. This command allows
%% the author to define a more concise list
%% of authors' names for this purpose.
\renewcommand{\shortauthors}{Benabdallah}

%%
%% The abstract is a short summary of the work to be presented in the
%% article.
\begin{abstract}

Explainable AI (XAI) interfaces seek to make large language models more transparent, yet explanation alone does not produce understanding. Explaining a system's behavior is not the same as being able to engage with it, to probe and interpret its operations through direct manipulation. This distinction matters for scientific disciplines in particular: scientists who increasingly rely on LLMs for reading, citing, and producing literature reviews have little means of directly engaging with how these models process and transform the texts they generate. 
In this ongoing design research project, I argue for a shift from explainability to interpretative engagement. This shift moves away from accounts of system behavior to instead enable users to manipulate a model's intermediate representations. Drawing on textual scholarship, computational poetics, and the history of reading and writing technologies, including practices such as marginalia, glosses, indices, and annotation systems, I propose interpretative interfaces as interactive environments in which non-expert users can intervene in the representational space of a language model. More specifically, such interfaces will allow users to select a token and follow its trajectory through the model's intermediate layers. This way, they can observe how its semantic position shifts as context is processed, and possibly annotate the transformations they find useful or meaningful. The same way readers can create their own maps within a book through annotations and bookmarks, interpretative interfaces will allow users to inscribe their reading of a model's internal representations. The goal of this project is to reframe AI interpretability as an interaction design project rather than a purely technical one, and to open a path toward AI-mediated reading that supports interpretative engagement and critical stewardship of scientific knowledge.

\end{abstract}

%%
%% The code below is generated by the tool at http://dl.acm.org/ccs.cfm.
%% Please copy and paste the code instead of the example below.
%%
%\begin{CCSXML}

%\end{CCSXML}

%\ccsdesc[500]{Do Not Use This Code~Generate the Correct Terms for Your Paper}
%\ccsdesc[300]{Do Not Use This Code~Generate the Correct Terms for Your Paper}
%\ccsdesc{Do Not Use This Code~Generate the Correct Terms for Your Paper}
%\ccsdesc[100]{Do Not Use This Code~Generate the Correct Terms for Your Paper}

%%
%% Keywords. The author(s) should pick words that accurately describe
%% the work being presented. Separate the keywords with commas.
%% A "teaser" image appears between the author and affiliation
%% information and the body of the document, and typically spans the
%% page.

%\begin{teaserfigure}
%  \includegraphics[width=\textwidth]{sampleteaser}
 % \caption{Seattle Mariners at Spring Training, 2010.}
 % \Description{Enjoying the baseball game from the third-base
 % seats. Ichiro Suzuki preparing to bat.}
%  \label{fig:teaser}
%\end{teaserfigure}

%\received{20 February 2007}
%\received[revised]{12 March 2009}
%\received[accepted]{5 June 2009}

%%
%% This command processes the author and affiliation and title
%% information and builds the first part of the formatted document.
\maketitle

\section{Introduction}
\subsection{The Limits of Explanations}
AI systems, particularly LLMs, are increasingly mediating the activities of knowledge production, such as reading, writing, citation, literature reviews, publishing, and the discovery of scholarly and scientific literature. LLMs are changing how researchers encounter text and, to some extent, how they assess their relevance and authority \cite{Whitfield2023ElicitALA, Kusumegi2025ScientificPIA, Musslick2024AutomatingTPA}. In the face of these shifts, there is a growing body of research concerned with developing explainable AI interfaces for end-users. Domains such as human-centered explainable AI (HCXAI), for instance, complement the goals of AI interpretability by moving beyond the objectives of algorithmic transparency alone and instead foreground the sociotechnical questions of who can understand and make sense of AI systems, and how \cite{hcxai}.
%AI systems, particularly LLMs, are increasingly mediating the activities of knowledge production, such as reading, writing, citation, literature reviews, publishing, and the discovery of scholarly and scientific literature. LLMs are changing how researchers encounter text and, to some extent, how they assess their relevance and authority \cite{Whitfield2023ElicitALA, Kusumegi2025ScientificPIA, Musslick2024AutomatingTPA}. In the face of these shifts, there is a growing body of research concerned with developing explainable AI interfaces for end-users. 
%\gb{There is probably a better way to talk about that.} 
%Domains such as human-centered explainable AI (HCXAI) \cite{hcxai}, for instance, complement the goals of AI interpretability by moving beyond the objectives of algorithmic transparency alone and instead foreground the sociotechnical questions of who can understand and make sense of AI systems, and how \cite{10.1145/3613904.3642474}.
Especially in the context of reading and writing, which inform attention, interpretation, and other cognitive processes \cite{ong_2012, wolf_2008}, ethical concerns emerge not only in model training but at the interface level, which structure how systems are understood and used. By being the sites where the modalities of interaction, and therefore of interpretation, meaning-making, and agency are negotiated, interfaces articulate the possibilities of participation 
%in customization, extension, and appropriation of 
in the activities of knowledge production. Interface design choices fundamentally shape how information is received, what counts as legitimate scholarly work, as well as what reading and writing practices are valued and sustained.
%, and who can participate meaningfully in knowledge production.

As AI tools become part of the infrastructures of academic work, mediating individual tasks and the social practices of citation and peer review, the interfaces through which scholars engage with these systems have to be considered as interventions in the knowledge commons. Designing interpretative interfaces, then, is about sustaining the interpretative practices that have historically been supported by certain reading and writing practices that invite readers and users into the information architecture of the system.

%AI interfaces used in academic and scholarly work are a particularly high-stakes domain as they shape epistemic practices and the technical conditions of collective knowledge formation.
%despite the necessity of developing explainable AI interfaces, 
Explainability is also only one aspect of the project of opening up AI systems to a wider range of interests and concerns. Many explainable AI interfaces focus on developing interactive modalities that explain the behavior or operations of LLMs and classifier systems. These modalities include graphs \cite{google}, dashboards \cite{dijk_2019}, chat-based interactions \cite{Kovari_2025}, and natural language explanations \cite{wang-etal-2024-llmcheckup}, among others. These explanations, while necessary, also present the paradoxical issue of adding an additional layer of abstraction on already complex and opaque systems. Additionally, explanations tend toward closure, resolution, and often assume authority; as such, they do not cultivate the inquisitive and critical attitude that characterize interpretation.
%and tend to flatten inquiry rather than sustain it. 

\subsection{Interpretation as Material Engagement}
Interpretation is defined as the act of explaining the meaning of something, but modern hermeneutics, the tradition dealing with interpretation, complicates what this process of arriving at an explanation is. In reader-response theory, Wolfgang Iser argued that meaning is not inherent to the text but a ``dynamic happening'' produced through the reader's active engagement with the text \cite[p.~22]{iser_1978}.
This is in part because the text, like all information systems, has gaps and indeterminacies that the reader resolves internally through creative and situated reconfigurations, ``working things out'' rather than receiving pre-formed explanations \cite[~p.276]{iser_1978}.
Reading, therefore, is a form of manipulation: the reader \textit{intervenes} in the material of the text, filling gaps, rearranging expectations, and reconfiguring meaning against the ``horizon'' of their own histories and concerns. Understanding does not happen as a transmission but as an act of construction, one that is enabled through hands-on engagement.

This insight extends beyond literary texts and applies to all information systems. Manipulation, in the sense of direct, material intervention in a system, is a more powerful vector of understanding than explanation. Explanations tend toward closure by resolving ambiguity, and assume an authoritative position that comes from the system (or its authors) rather than the users. Interpretation by contrast sustains inquiry. In fact, interpretation thrives on resistance. As the literary scholar Yves Citton writes,

\begin{quote}
``[a] century of interpretive practice has taught us to turn sites of resistance to meaning into strategic points for the emergence of ``truth.'' Nothing is more tedious than a textual analysis that merely confirms how clear and well-ordered the narrative is ... . Interpretation truly begins with the grain of sand that jams the machine that produces obviousness, ... with the detection of an incoherence, an unexpected detail, a ``deviation''.'' \cite[p.~128]{Citton_lire} (my translation).
\end{quote}

Ambiguous, polysemic, and even ``glitchy'' outputs are productive sites of engagement; opportunities to question and re-imagine systems rather than merely optimize them. Understanding emerges precisely where the system resists easy explanation.
Explanation-centric interfaces in AI place epistemic authority at the system level: the system explains itself, and the user receives it as the voice of authority. This has the paradoxical effect of reducing agency and accountability; many explainable AI interfaces tend to substitute explanation for the ability to probe and materially engage with a system. By doing so, these interfaces foreclose the possibility of inquiry that they could instead support. Additionally, they normalize (or unproblematize) the epistemic value of transparency, and subtly relocate authority at the system level. Interpretative interfaces, by contrast, give users the means to manipulate language models not just at the interface level but at the operational level by giving access to the intermediate layers. They allow intervention in the information architecture of a system, and enable users to leave traces of their interpretative work, thereby constructing understanding on their own terms.

\section{The Book: A Historical Model for Interpretative Interface Design}
The history of the book is a strong example of an information system that became gradually easier to manipulate (and interpret). Through the development of both technical innovations and social practices over centuries, the book evolved to allow more readers to intervene in it. This history can be instructive for the design of AI interfaces.

For centuries, codices were clunky and heavy artifacts that relied on specialized information architectures and conventions, such as the absence of space between words or the use of Merovingian and Visigothic scripts, which were often illegible to most but a few domain experts (monks and scribes). The literal and technical heftiness of early manuscripts, as well as their interface design, meant that only a few could produce and use books \cite{Levy_1993}. Books were, effectively, `expert systems' \cite{suchman_2007}. Over the centuries, a series of technical developments made the book progressively more usable. Such developments include the normalization of the caroline minuscule as a calligraphic standard, the refinement of paper making techniques so that it became smoother, new binding and folding techniques which shrunk the size (and reduced weight) of codices, and eventually the printing press, which, as Elizabeth Eisenstein argued, radically altered the conditions under which texts circulated and endured \cite{Eisenstein_2012}.

But usability alone does not account for how the book ``opened up'' to interpretation. What proved equally decisive was the development of spaces for intervention: margins, interlinear gaps, and the social practices they enabled, such as glosses, marginalia, commonplace books, and later, independent letterpress shops and zines. These graphic and social infrastructures, now largely taken for granted, offered the material conditions for interpretative practice. The margin is, in a sense, Iser's ``gap'' made physical: a space within the system where the reader can intervene, annotate, contest, and construct meaning. The book became what Ivan Illich called a ``convivial'' tool \cite{illich_1985} because it invited participation at the level of its own information architecture and, through doing so, extended the technology of the book to a wider range of interests and concerns.

The takeaway for the design of XAI interfaces is that accessibility and interpretability are not best supported by explanation or simplification but by enabling what is simultaneously a material and symbolic form of participation, which is interpretation.

\section{Towards Interpretative Interfaces for LLMs}
LLMs, like books, have material affordances that can be leveraged to support interpretation, and without asking users to master technical explanations. Just as the margins and interlinear gaps of the page functioned as sites for annotation and interpretation, LLMs contain representational spaces that shape how meaning is internally produced: the intermediate layers of transformer-based models. While not immediately accessible to end-users, techniques from the field of AI interpretability have made it possible to inspect and manipulate these layers \cite{skean2025layerlayeruncoveringhidden, skean2024doesrepresentationmatterexploring, Kowalska2025UnboxingTBA, Raukur2022TowardTAA, Linardatos2020ExplainableAAA}. While these techniques have predominantly been used in technical scenarios to better understand how models work (and, by extension, make them better), I suggest that these intermediate representations can also be approached as part of the interaction itself.

I propose \textit{interpretative interfaces}: interface designs that allow users to interact with and annotate the intermediate representations of LLMs. Instead of confining interaction to the input-output level, interpretative interfaces enable direct manipulation of a model's internal operations. The interaction is structured around \textit{token-centric reading}: rather than treating the model as a black box to be explained through additional layers of abstraction (through graphs, plots, and dashboards), users select individual input tokens and trace their transformations through the model's layers, observing how semantic positions shift as context is processed. This reframes the interaction from model-centric interpretability (what did the system do?) to token-centric reading (how was this particular unit transformed, and how did it influence ``meaning''). The goal is not to exhaustively expose a model's operations, which would be overwhelming, but to create partial windows (like margins in a manuscript) through which users can engage with, annotate, and begin interpreting how meaning is constructed within the system.

This approach also raises questions about how intermediate representations are themselves visualized. As media theorist Johanna Drucker has argued, dominant modes of graphic display tend to be treated as transparent representations of the phenomena they point to, when in fact they are ``as unsuited to modeling interpretation as a thermometer is for measuring the warmth of a human emotion or the strength of an embrace'' \cite[~p.5]{drucker_2020}. Many phenomena, including language, exceed the bounds of representation--the map is not the territory \cite{korzybski_1933}. Representations, whether expressed through maps, visualizations, or interfaces, are always already interpretations. Rather than assuming transparency, interfaces that honor this fact need to call attention to their own materiality, their own particular way of representing the world, so that representation does not become something to be wary of, but something to work with. Interfaces optimized solely for clarity and efficiency may actively suppress the kinds of interpretive inquiry through which users could develop critical understanding of AI-mediated knowledge production. Interpretative interfaces should instead foreground the constructedness of their own representations, offering an alternative to the dominant chat-based paradigm, whose anthropomorphic framing risks obscuring the material and contingent nature of model operations.

\section{Designing Interpretative Interfaces}
Following a two-month ethnographic study of AI-mediated reading practices in scientific disciplines, this project is currently in an exploratory design phase. Our approach is organized around three interrelated design spaces: \textit{action} (what users can do), \textit{representation} (how intermediate layers are visualized), and \textit{interaction} (how users physically manipulate the interface).

\subsection{Action: Follow the Token}
The core action of an interpretative interface is to \textit{follow a token}. A user selects a token, from an input prompt or a model's output, and traces its embedding trajectory through the model's intermediate layers, observing how its position in semantic space shifts as context is processed. These transformations are visualized through dimensionality reduction techniques (such as PCA or UMAP), and can be projected into a 2D visualization where users can see, compare, and annotate the trajectories of multiple tokens simultaneously. The result is a form of token-centric reading, where users directly engage with how meaning is produced at the smallest unit (token), layer by layer.

This interaction is designed to enable what we call a ``glossed model,'' drawing on the tradition of medieval manuscript glosses, which were annotations, commentaries, and translations added between lines or in margins to interpret and educate. Similarly, interpretative interfaces allow users to annotate specific token-layer combinations, creating a cumulative record of the traces of their interpretative work with the model's operations.

\subsection{Representation: A Swatchbook of Intermediate Layers}
Since layer representation for non-experts is an important yet underexplored design space, we are developing a ``swatchbook'' of representation modes. Representation modes include panels, strata, organic shapes, constellations, grids, rays, mind maps, and translucent overlays, among others. Each mode supports a different type of sense-making of how layers ``behave.'' The swatchbook is itself a design research contribution that offers a design vocabulary for intermediate layers of transformer-based models. By exploring these different representation modes, the swatchbook emphasizes that there is no single ``correct'' way to see a model's intermediate operations, but instead that the choice of representation shapes to a significant degree what can be noticed, compared, and therefore understood and questioned.

\subsection{Interaction: Manipulation as a Modality}
Interpretative interfaces emphasize physical, tangible interaction metaphors, such as dragging layers with ``hand-cursors,'' grabbing tokens from a ``token bag,'' scrubbing through layers as one might scrub through a video timeline, capturing ``stills'' as they go. These interactions aim to foreground manipulation as a modality of understanding, therefore complementing more traditional XAI initiatives. 

\subsection{Implementation and Next Steps}
The current prototype uses TransformerLens \cite{nanda_2023} to inspect a small transformer model (GPT-2, 124M parameters, 12 layers), with a Python backend serving embedding data to a browser-based UI. We are currently conducting the design work on two parallel tracks: the first is to work out the interaction flows and representation modes in Figma; the second is to develop the technical backend to extract and reduce embedding trajectories so that they can be visualized. These tracks will ultimately converge in a functional prototype that connects model data to the designed interactions.

Following this exploratory phase, we will host participatory design workshops at the University of Washington, bringing together scientists, educators, artists, and researchers to test the early prototypes and iterate on their design. We wish to ground discussions around the possibility of supporting critical and interpretative engagement with AI tools via the prototypes themselves. They will be used as ``objects to think with'' \cite{papert_1993}, so as to scaffold conversations about what LLM interfaces might be. An important goal of this project is to open up the sociotechnical imaginaries \cite{jasanoff_kim_2015} of how people interact with language models. 

\section{Conclusion}
This paper has argued that interpretation, not explanation, is the primary means through which people develop understanding of complex systems, and that interpretation requires the material and conceptual means to manipulate. In the context of AI for science, this is crucial as scientists increasingly rely on AI-augmented tools to find, read, and cite literature, and therefore build understandings of their field and research space.

Interpretative interfaces contribute to the space of explainable AI (XAI) not through explanation but through epistemic actions \cite{Kirsh1994-KIRODE}, specifically the manipulation and annotation of intermediate layers. By structuring interaction around token-centric reading (selecting tokens, tracing their trajectories through layers, annotating the transformations that lead to semantic relevance) they offer users a way to engage with LLMs that is fundamentally different from receiving an explanation. 

As LLMs increasingly mediate the practices of knowledge production, such as reading, writing, citation, literature review, and publishing, the interfaces through which scholars and scientists engage with these systems become part of the knowledge infrastructure itself. We cannot risk closing the possibilities of critical inquiry and material intervention that would foster the kind of interpretative engagement that is crucial for healthy knowledge ecologies. Designing for interpretation sustains the conditions for critical inquiry and the kind of material participation that has historically sustained the knowledge commons. Open scholarship depends not only on open access and open data but on open tools, whose architecture invites intervention and manipulation. Interpretative interfaces are a step toward that goal.

\bibliographystyle{ACM-Reference-Format}
\bibliography{software}

\end{document}